\documentclass[conference,dvipsnames]{IEEEtran}
\IEEEoverridecommandlockouts
\usepackage{cite}
\usepackage{amsmath,amssymb,amsfonts}
\usepackage{algorithmic}
\usepackage{graphicx}
\usepackage{eurosym}
\usepackage{textcomp}
\usepackage{xcolor}
\usepackage{xargs}
\usepackage{color,listings}
\usepackage{tabularx}
\usepackage{tabularray}
\usepackage{listings}
\usepackage{caption}
\usepackage{multirow}
\usepackage{ragged2e}
\usepackage[colorinlistoftodos,prependcaption,textsize=scriptsize]{todonotes}
\def\BibTeX{{\rm B\kern-.05em{\sc i\kern-.025em b}\kern-.08em
    T\kern-.1667em\lower.7ex\hbox{E}\kern-.125emX}}

\newcommandx{\john}[2][1=]{\todo[linecolor=Blue,backgroundcolor=Blue!25,bordercolor=Blue,#1]{\textbf{John comments: }#2}}
\newcommandx{\mitra}[2][1=]{\todo[linecolor=OliveGreen,backgroundcolor=OliveGreen!25,bordercolor=OliveGreen,#1]{\textbf{Mitra comments: }#2}}
\newcommandx{\pragyan}[2][1=]{\todo[linecolor=Red,backgroundcolor=Yellow!25,bordercolor=Red,#1]{\textbf{Pragyan comments: }#2}}
\newcommandx{\tdb}[2][1=]{\todo[linecolor=Mulberry,backgroundcolor=Orchid!25,bordercolor=Mulberry,#1]{\textbf{Travis comments: }#2}}
\newcommandx{\rambod}[2][1=]{\todo[linecolor=Bittersweet,backgroundcolor=Bittersweet!25,bordercolor=Bittersweet,#1]{\textbf{Rambod comments: }#2}}
\newcommandx{\thomas}[2][1=]{\todo[linecolor=RoyalBlue,backgroundcolor=ProcessBlue!25,bordercolor=RoyalBlue,#1]{\textbf{Thomas comments: }#2}}

\newcommandx{\citeme}[2][1=]{\todo[linecolor=red,backgroundcolor=red!25,bordercolor=red,#1]{\textbf{CITE} #2}}
\newcommandx{\fillin}[2][1=]{\todo[linecolor=red,backgroundcolor=red!25,bordercolor=red,#1]{\textbf{FILL IN} #2}}

\begin{document}

\title{Demystifying Feature Requests: Leveraging LLMs to Refine Feature Requests in Open-Source Software}

\author{\IEEEauthorblockN{Pragyan K C\textsuperscript{1}, Rambod Ghandiparsi\textsuperscript{1}, Thomas Herron\textsuperscript{1}, John Heaps\textsuperscript{1}, Mitra Bokaei Hosseini\textsuperscript{1}}
\IEEEauthorblockA{\textsuperscript{1}University of Texas at San Antonio, San Antonio, TX, USA\\
\textit{
[pragyan.kc, rambod.ghandiparsi, thomas.herron, john.heaps, mitra.bokaeihosseini]@utsa.edu, }
}

}

\maketitle
\begin{abstract}

The growing popularity and widespread use of software applications (apps) across various domains have driven rapid industry growth. Along with this growth, fast-paced market changes have led to constantly evolving software requirements. Such requirements are often grounded in feature requests and enhancement suggestions, typically provided by users in natural language (NL). However, these requests often suffer from defects such as ambiguity and incompleteness, making them challenging to interpret. Traditional validation methods (e.g., interviews and workshops) help clarify such defects but are impractical in decentralized environments like open-source software (OSS), where change requests originate from diverse users on platforms like GitHub. 
This paper proposes a novel approach leveraging Large Language Models (LLMs) to detect and refine NL defects in feature requests. Our approach automates the identification of ambiguous and incomplete requests and generates clarification questions (CQs) to enhance their usefulness for developers. To evaluate its effectiveness, we apply our method to real-world OSS feature requests and compare its performance against human annotations. In addition, we conduct interviews with GitHub developers to gain deeper insights into their perceptions of NL defects, the strategies they use to address these defects, and the impact of defects on downstream software engineering (SE) tasks.

\end{abstract}

\section{Introduction}


As software continuously evolves to meet users' growing and changing needs, the ability to efficiently accommodate new features and enhancements has become a crucial factor in maintaining competitiveness and ensuring user satisfaction. Feature requests serve as an essential mechanism for guiding software evolution, as they allow users to propose new functionalities or improvements based on their experiences~\cite{carreno2013analysis,oriol2018fame,dalpiaz2019re,di2016would,maalej2015bug,yang2015identification}. These requests originate from diverse sources, such as application (app) store reviews, issue-tracking platforms like GitHub, and dedicated user feedback forums~\cite{oriol2018fame}. While feature requests offer valuable insights into user needs, they are written in natural language (NL), making them susceptible to defects such as \textbf{ambiguity} and \textbf{incompleteness}~\cite{handbook2003contract}. These defects arise due to communication errors, missing information, or lack of technical expertise on the requester's part. As a result, developers must interpret and refine feature requests, a process that can lead to incorrect assumptions, flawed implementations, and ultimately, reduced software quality~\cite{zowghi2003interplay}. The cost of identifying and mitigating NL defects increases as software development progresses~\cite{FWK+17,BB01}, making early detection and validation of feature requests essential.

Traditionally, change requests, including feature requests, are refined through iterative elicitation and negotiation techniques such as interviews, prototyping, and workshops~\cite{debnath2021ideas, ribeiro2020prevalence}. While these methods work effectively in closed-source software development, where stakeholders are readily accessible~\cite{oriol2018fame}, they are impractical in open-source software (OSS) development~\cite{mockus2002two,kuriakose2015open}. Unlike traditional software organizations, OSS development is decentralized, driven by a globally distributed community of developers and contributors who voluntarily engage in projects for various motivations, including the need for specific functionalities, skill development, career advancement, and personal interest~\cite{crowston2008free,fellerframework,shah2006motivation}. In OSS, feature requests are often submitted by users who may not have direct access to developers, making it difficult to validate or refine requests through conventional methods. Consequently, OSS developers, who oversee and manage development activities~\cite{crowston2005social}, must process incomplete or ambiguous requests without direct clarification from requesters, leading to potential misunderstandings and suboptimal implementations.


This paper addresses the following challenges in OSS development: (1) Ensuring the quality of feature requests by detecting NL defects; and (2) Supporting OSS developers in validating and refining feature requests in the absence of direct interaction with requesters. To tackle these challenges, we propose a practical framework that leverages Large Language Models (LLMs). 
Our study focuses specifically on feature requests submitted through GitHub issues, which serve as a primary mechanism for planning, discussing, and tracking development work in OSS repositories. 


This paper makes three key contributions, each addressing a distinct aspect of the aforementioned challenges. 
(1) We conduct an empirical analysis of feature requests to identify common NL defects, including ambiguity and incompleteness in feature requests and develop automated methods leveraging LLMs to detect and classify NL defects in feature requests. We evaluate our defect detection methods using feature requests from Mastodon and Signal, comparing their effectiveness against human annotations and analyzing their potential to enhance OSS development workflows. (2) We develop methods to automatically generate clarifying questions (CQs) to address ambiguity and incompleteness while preserving the requester’s original intent~\cite{potts1994inquiry}. Further, we evaluate the effectiveness of generated CQs by assessing the quality and relevance of the generated CQs in improving feature request clarity. 
(3) We conduct interviews with Signal developers to explore how developers perceive NL defects in feature requests, how they address these defects in practice, and the effects of NL defects on downstream software engineering (SE) tasks, including design, implementation, testing, and maintenance. 
Our contributions are publicly available in our open-source repository~\cite{AnonymizedRepo2025}.

The remainder of this paper is organized as follows: Section~\ref{sec:background} presents background and related work; Sections~\ref{sec:approach} and ~\ref{sec:experimentDesign} entail the approach and experiment designs; Sections~\ref{sec:results} and ~\ref{sec:discussion} entail results and discussion; Sections~\ref{sec:threats_validity} and ~\ref{sec:conclusion} contain threats to validity and conclusion.

\section{Background and Related Work}\label{sec:background}


\subsection{Large Language Models (LLMs)}
LLMs have rapidly emerged as incredibly powerful resources on a wide range of tasks and domains, from translation and summarization to more complex tasks like medical diagnosis and legal review~\cite{vaswani2017attention,chang2024survey,zhao2023survey,devlin2019bert}. Their ability to comprehend and generate human-like text has changed many industries and made LLMs very attractive to undertake complex problems in SE~\cite{hou2024large}. 
Prompt engineering is the deliberate combination 
of various elements, such as persona, task-specific instructions, and output constraints, to create prompts that direct LLMs toward generating high-quality, task-related responses~\cite{liu2023pre,radford2019language,vatsal2024survey,ouyang2022training}. 
In this section, we review two prompting strategies. 
\textbf{In-context Learning (ICL)} exploits the LLM's intrinsic ability to learn from examples provided explicitly in the prompt. In the \textbf{few-shot learning (FSL)} approach, a few example input-output pairs are given to show both the task and the desired output style~\cite{brown2020language,dong2022survey,perez2021true}. Also, the employment of positive and negative examples helps in refining the outputs of the model by imposing the desired characteristics and minimizing potential biases or errors~\cite{mo2024c,nguyen2023context,santos2024requirements}. 

\subsection{Natural Language (NL) Defects in Requirements}

NL enables communication between the requester, software analyst, developer, other users, and stakeholders who may have different backgrounds, often
having little or no additional training in software development~\cite{pohl1996requirements}. However, NL requests are nonetheless prone to defects, including ambiguity and incompleteness~\cite{handbook2003contract,ezzini2021using,pohl1996requirements}. 
Ambiguity occurs when language used in requirements can be interpreted in more than one way~\cite{berry2004ambiguity}. Incomplete feature requests are those that lack crucial information needed to fully describe the intended system behavior~\cite{zowghi2003interplay}. According to Boehm~\cite{boehm1984verifying}, a complete specification must detail all aspects required for correct system functionality. Both defects create challenges, hindering validation and verification processes, and ultimately leading to bad communication between developers and clients~\cite{zowghi2003interplay}.


Ambiguity can be categorized into: lexical ambiguity, where a word has multiple meanings; syntactic ambiguity, where sentence structure allows multiple interpretations; semantic ambiguity, where meaning is unclear due to word relationships; pragmatic ambiguity, where interpretation depends on context; and vagueness, where a term is subjective or imprecise~\cite{handbook2003contract}.

In a conventional industrial setting, ambiguity in requirements is rarely problematic because it is actively discussed and resolved through goal-oriented conversations between stakeholders~\cite{ribeiro2020prevalence,de2010ambiguity,philippo2013requirement}. 
However, this level of refinement is difficult to achieve in OSS, where requesters and developers often lack direct communication channels. Although platforms like GitHub facilitate open discussions through comments, these conversations can become disorganized, unfocused, and cluttered with unrelated topics, making it challenging to reach a clear, actionable understanding of the request~\cite{heck2017framework}. Furthermore, discussions may receive irrelevant input from other users, further obscuring the original intent of the feature request and hindering effective implementation~\cite{heck2017framework}. 

Prior research in requirements engineering (RE) and SE has tried to detect NL defects including ambiguity and incompleteness~\cite{fantechi2023rule, ezzini2022automated, zait2018addressing, femmer2017rapid, ferrari2018detecting, yang2010automatic, gleich2010ambiguity, mu2020nero, seki2019detecting, arora2019empirical, luitel2024improving, lian2024reqcompletion}. 
Ambiguity detection studies lack comprehensive categorization, focusing on narrow classifications, such as lexical or pragmatic ambiguity~\cite{fantechi2023rule, zait2018addressing, femmer2017rapid, ferrari2018detecting, yang2010automatic, gleich2010ambiguity, mu2020nero, seki2019detecting}. Incompleteness studies often rely on probabilistic models to predict the likely terms missing from sentences based on statistical patterns in the data~\cite{luitel2024improving}. However, by doing so, they can inadvertently misinterpret the requester's original intent. 
In addition, existing research focuses primarily on the detection of NL defects rather than their clarification and refinement~\cite{fantechi2023rule, ezzini2022automated, zait2018addressing, femmer2017rapid, ferrari2018detecting, yang2010automatic, gleich2010ambiguity, mu2020nero, seki2019detecting, arora2019empirical, luitel2024improving}. While some studies attempt to resolve defects, their scope is often limited to specific patterns and domain models, addressing only certain types of missing information ~\cite{veizaga2024automated,arora2019empirical,ezzini2022automated}. 

\subsection{Open-Source Software (OSS) and GitHub}
OSS development is distinguished by its decentralization and volunteerism, which substitutes for traditional and formal requirements engineering practices with informal, ad hoc methods~\cite{10.1145/2089125.2089127,heck2013analysis}. Developers generally use personal knowledge and community forums to create and evolve their requirements, leading to less-structured practices than are typical of conventional closed-source environments~\cite{felfernig2018towards,7431307}. 
Differing agendas and communication styles directly impact how requirements are elicited, negotiated, and prioritized~\cite{6401114}.

Researchers have developed various strategies, including automated classification systems, standardized labeling, and iterative discussion protocols, to streamline issue management~\cite{chaparro2019using,mahmud2025combining} for GitHub, as a leading platform for OSS development. 
Different studies have enhanced our understanding of technical approaches (e.g., automated issue classification) and process-oriented methods (e.g., standardized labeling, iterative discussions), typically involving AI and ML based tools that generate predefined suggestions based on patterns and historical data, to improve efficiency in software development practices~\cite{anvik2006should,bettenburg2008makes}, but they often overlook the inherent ambiguity and incompleteness that can affect issues. 

\vspace{-0.5em}
\section{Approach}\label{sec:approach}

\begin{figure}
\vspace{-1em}
	\centering 
        \includegraphics[width=0.47\textwidth]{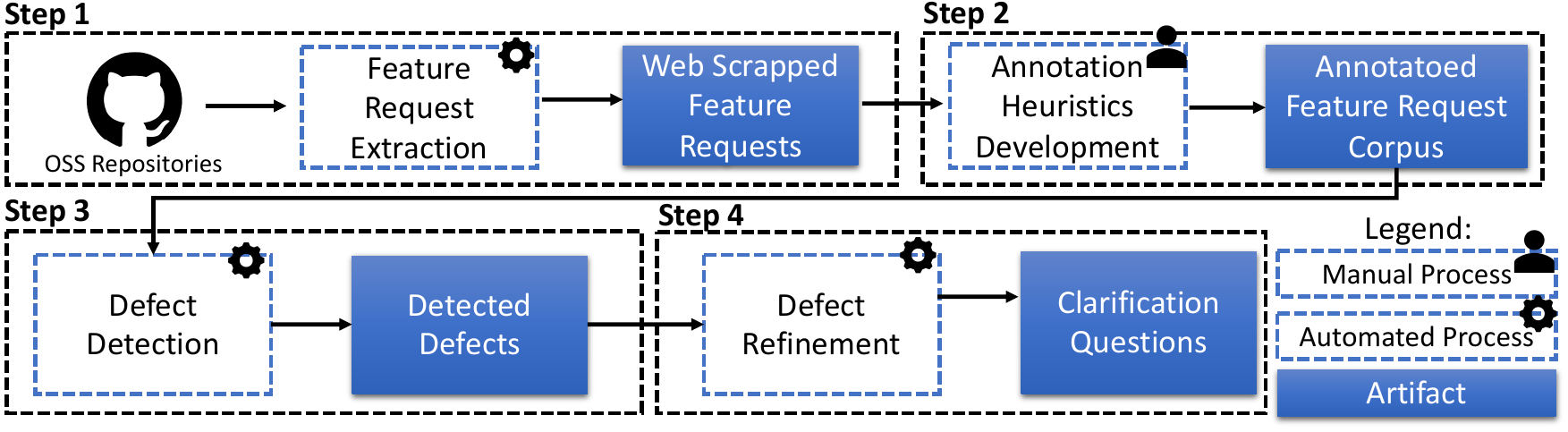}
	\centering
	\caption{Approach Overview}
	\label{fig:approach-overview}
\vspace{-2em}
\end{figure} 

In this section, we describe our approach for detecting and clarifying defects in feature requests. Figure~\ref{fig:approach-overview} shows four different steps involved in our approach. In the first step, we create a dataset by collecting feature requests from open-source GitHub projects; in the second step, we establish a ground truth (GT) for identifying and clarifying defects in the dataset through human annotations. In the third step, we utilize LLMs to detect NL defects. Finally, in step 4, we generate CQs to refine feature requests with defects. 

\subsection{Step 1: Feature Requests Collection}\label{sec:originalDataset}
The goal of Step 1 is to create a feature request dataset. 
To this end, we select Mastodon\footnote{https://github.com/mastodon/mastodon-android} and Signal\footnote{https://github.com/signalapp/Signal-Android}, two well-known OSS projects on GitHub that serve as open-source alternatives to commercial platforms. These projects focus on messaging and social media, ensuring a user interface (UI) is present, which is an important factor in capturing user-driven feature requests. 
We capture a broader range of user contributions, including non-technical users, unlike prior studies~\cite{jimenez2023swe} that focused on expert developers in Python repositories. 

To collect feature requests from these GitHub repositories, we use BeautifulSoup~\footnote{https://pypi.org/project/beautifulsoup4/\#description}, an HTML scrapper. We navigate to the issue section of these repositories and filter the issues labeled `Feature Request' or `Feature'. Next, we extract the issue numbers associated with the `Feature Request' or `Feature' label. Then, we run the web scraper on each of these issue numbers to collect detailed information, including the request's title, description, author, date, and discussion. In total, we collect a dataset of 476 feature requests, including both open and closed issues, from the two repositories~\cite{AnonymizedRepo2025}. 
Our preliminary analysis reveals that these feature requests in Mastodon and Signal experience long turnaround times, averaging 254 days and 599 days, respectively, from submission to completion. This suggests challenges in managing feature requests, potentially leading to delays in addressing user needs.

\subsection{Step 2: Feature Request Annotation}

\begin{figure}
\vspace{-2em}
	\centering 
	\includegraphics[width=0.45\textwidth]{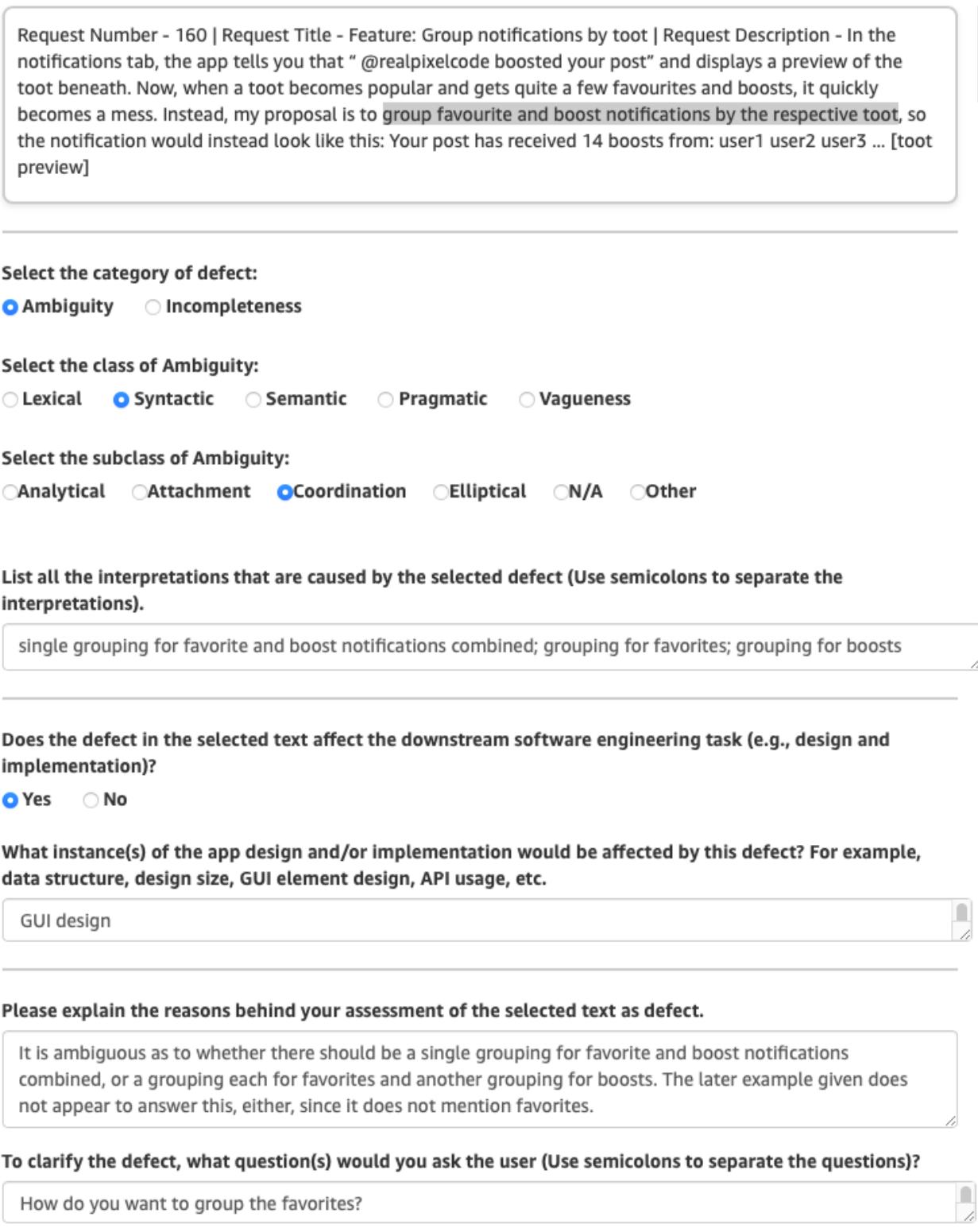}
	\centering
	\caption{Annotation Tool}
	\label{fig:annotation}
\vspace{-2em}
\end{figure}

The goal of Step 2 is to create an annotated feature request corpus. For this reason, we first randomly select 15 feature requests from the dataset and analyze them using grounded theory~\cite{corbin2008basics}. Through this analysis, we identify two broad classes of NL defects (i.e., ambiguity and incompleteness). As an example, the top part of Figure~\ref{fig:annotation} illustrates Feature Request Number 160, from Mastodon's GitHub repository. 
In Mastodon, ``favorites'' function as likes for posts (known as toots), and ``boosts'' are equivalent to retweets on Twitter. The request contains several NL defects, including unclear grouping of notifications (boosts vs. favorites), an undefined layout for grouped notifications, and vague language, such as describing notifications as ``messy'' when a toot is popular. There is no clear limit for the number of names displayed. Although 14 boosts are mentioned, further refinement is needed to define a display limit and determine what happens when the limit is reached. 


We also analyze the 15 requests to determine whether any of their comments addressed the NL defects identified during our annotations~\cite{AnonymizedRepo2025}. We observe mentions of difficulty in implementing the feature request, expressing the usefulness of the requested feature in solving an issue, requesting additional features, reporting bugs, or referring to another request by citing its request number (unique ID on the repository). Notably, seven of the feature requests had no follow-up comments at all. 
Still, 8 of the 15 feature requests have comments, indicating that GitHub issue management has the potential for community engagement to refine and clarify feature requests.  
However, this process is not always consistent, as evidenced by the feature requests that have no follow-up comments or irrelevant comments that do not help with clarifying the requests. 


To ensure consistency with prior work, we adopt ambiguity sub-classes from Berry et al.~\cite{handbook2003contract}. Based on our analysis, we develop an annotation tool for labeling NL defects (ambiguity and incompleteness), sub-classes (lexical, syntactic, semantic, pragmatic, and vagueness), interpretations, reasoning, and CQs\cite{AnonymizedRepo2025}. Figure~\ref{fig:annotation} illustrates the tool. Initially, the authors familiarize themselves with handbook definitions, independently analyze 15 feature requests, and meet to discuss findings. This process results in a heuristics document outlining the annotation procedures (see~\cite{AnonymizedRepo2025}). 

We then select 100 feature requests from the dataset through stratified sampling, using the request status (open with discussion or closed with discussion) as the stratification criterion. This approach ensures that we capture both completed requests and those with clarifications about potential issues. The 100 selected requests are subsequently divided into three annotation batches of 20, 50, and 30. This phased approach allows us to start with a smaller batch to build familiarity and consistency in the annotation process before scaling up to larger sets.

Two authors perform the annotations: Annotator A, a fluent non-native English speaker with an extensive background in SE, and Annotator B, a native English speaker with a background in linguistics and cybersecurity. Annotators performed analysis concurrently but independently of one another, meeting weekly to discuss whether any issues were encountered that required clarifying or modifying the heuristics document. 

During annotation, if an ambiguity is detected, annotators highlight the relevant text segment and then specify sub-class of the ambiguity and offer their interpretations—since ambiguity requires at least two plausible interpretations~\cite{handbook2003contract}. For incompleteness, the annotators do not highlight text but instead describe what is missing from the feature request. Next, they assess whether the defect (i.e., ambiguity or incompletness) could impact any downstream SE task, explaining how it might do so, and provide reasoning for their assessment. 
The annotators then provide CQs to help clarify the defect and refine the feature request.

After annotating each batch, we reconcile annotations and calculate Cohen's Kappa~\cite{cohen1960coefficient} at three levels: (1) Defect-Level Ambiguity Kappa (identifying ambiguous tokens): 0.55, 0.65, and 0.65 across batches; (2) Sub-Class Ambiguity Kappa (agreement on specific ambiguity sub-classes): 0.54, 0.64, and 0.64; (3) Incompleteness Kappa (identifying missing information in feature requests): 0.65, 0.40, and 0.87. Kappa results indicate moderate agreement between annotators. On average, both annotators spend 15 to 20 minutes per feature request. Among the 100 requests, 25 contain no defects. Further details on annotation disagreements are available in our repository~\cite{AnonymizedRepo2025}, and a summary of the reconciled corpus is provided in Table~\ref{tab:statsForDefectsInGT}.



\begin{table}[]
\centering
\caption{Defect Numbers}
\label{tab:statsForDefectsInGT}
\small
\begin{tabular}{|l|c|c|c|c|c|c|}
\hline
 & \textbf{Lex.} & \textbf{Syn.} & \textbf{Sem.} & \textbf{Prag.} & \textbf{Vague.} & \textbf{Incomp.} \\ \hline
\textbf{\# Requests} & 24 & 16 & 9 & 17 & 11 & 57 \\ \hline
\textbf{\# Instances} & 42 & 24 & 10 & 32 & 14 & 57 \\ \hline
\end{tabular}
\vspace{-2em}
\end{table}

\vspace{-0.65em}
\subsection{Step 3: Defect Detection} 
The goal of Step 3 is to detect ambiguity and incompleteness using LLMs. Current LLMs face challenges in this task due to their lack of specialized training for handling ambiguous or incomplete statements, as well as varying levels of domain knowledge \cite{ferrari2024model}. Additionally, the manual annotation process is time-intensive, resulting in a relatively small corpus of only 100 annotated feature requests. To address these limitations, we apply in-context learning (ICL) under three settings: zero-shot, where the model receives no examples; few-shot learning (FSL), where the model is provided with a small number of labeled examples; and FSL with reasoning, which extends FSL by including the reasoning behind each label to help the model better understand the decision process. These settings are used to align the LLM with our detailed annotated corpus (see Section~\ref{subsec:Experiment1-DefectDetection}).

\vspace{-0.5em}
\subsection{Step 4: Defect Refinement}


The goal of Step 4 is to generate CQs to refine detected NL defects. LLMs excel at question-answering tasks when datasets contain well-defined questions with sufficient information for a unique answer~\cite{chung2024scaling, hoffmann2022training, kuhn2022clam}. However, studies show these models often hallucinate when processing text with NL defects~\cite{ji2023survey}. We explore ICL to generate CQs (see Section\ref{sec:Experiment2-DefectRefinement}).

\section{Experiment Design}\label{sec:experimentDesign}
We aim to address the following research questions. To achieve this, we design and conduct three key experiments.

\noindent\textbf{(RQ1)} To what extent can pre-trained LLMs detect and classify ambiguity and incompleteness in feature requests? 

\noindent\textbf{(RQ2)} What prompting methods improve the performance of pre-trained LLMs on detecting ambiguity and incompleteness? 

\noindent\textbf{(RQ3)} How can we generate clarifying questions to refine the detected ambiguity and incompleteness in feature requests?

\noindent\textbf{(RQ4)} How do LLM and human-generated CQs compare?

\noindent\textbf{(RQ5)} What are developers' perspectives on the NL defects in feature requests?





\subsection{Experiment 1: Defect Detection}\label{subsec:Experiment1-DefectDetection}

To address RQ1 and RQ2, we conduct experiments leveraging our annotated feature request corpus. We assess the performance of ICL using GPT-4o. 

To systematically structure the dataset for ICL, we create positive and negative sample sets for each defect class and sub-class. Given that we have five sub-classes of ambiguity and one class for incompleteness, we create separate positive-negative sets for each of these six types of defects. A positive sample for a defect corresponds to a feature request only if it contains annotations for the specified defect - either one specific sub-class of ambiguity or incompleteness, while a negative sample represents a feature request without that defect. Since a feature request can contain multiple defects, it may appear in multiple positive sets, but each sample in a positive set is isolated to the specific defect type.  

Next, we split each positive and negative set into training and testing subsets using a 30/70 ratio. This balance ensures a sufficient number of training examples for an FSL while maximizing the size of the test set. To this end, we obtain four data splits: \texttt{positiveTrain}, \texttt{negativeTrain}, \texttt{positiveTest}, and \texttt{negativeTest} for each ambiguity sub-class and incompleteness class. We concatenate \texttt{positiveTest} and \texttt{negativeTest} to create a single complete test set - \texttt{testingSet}. We create the splits with three different random seed values to maintain variability in our experiments. 

To construct FSL examples, for any experiment setting $E$ we randomly select $n$ samples from \texttt{positiveTrain} and \texttt{negativeTrain}, forming two sets: $s^+_n \subseteq \texttt{positiveTrain}$, $s^-_n \subseteq \texttt{negativeTrain}$. By incorporating both positive and negative examples, we improve the model’s ability to differentiate between defective and non-defective cases. Therefore, the total number of shots for the experiment $E$ is $2\times n$. We then apply a pair-based approach, where each positive sample is uniquely paired with a negative sample from $s^+_n$ and $s^-_n$, respectively. Unlike a full Cartesian pairing, where each positive sample is matched with all negative samples, we enforce a one-to-one pairing constraint: \textit{once a sample (positive or negative) is used in a pair, it cannot be reused in another pair}. Thus, the final paired set for experiment setting $E$ is shown as $E_n = \{ (p_i, q_i) \mid p_i \in s^+_n, q_i \in s^-_n,\text{each } p_i \text{ and } q_i \text{ used only once} \}$.

 Furthermore, for each experiment setting $E$, we also consider the permutation of pairs, ensuring that different pair orderings are incorporated in the training process. Given that $E$ consists of $n$ pairs, the total number of valid orderings is given by $P_n = n!$, where \( P_n \) represents the number of permutations of pairs in $E$.

\noindent\textbf{Experiment 1.1- Ambiguity Detection using FSL (Without Reasoning):} 
We detect and extract the text segments from the feature requests that contain a specific sub-class of ambiguity using two settings: Zero-shot and FSL. 
Listing~\ref{lst:1.1-AmbiguityDetection-FSL-Prompt} shows the prompt for the zero-shot setting, including the definition of the ambiguity sub-class, persona, instruction, and a feature request from the \texttt{testingSet} for detection. This setting does not include demonstration instruction and examples. 
For FSL, we provide examples as demonstrations in the prompt, where each example consists of a positive and a negative sample in a pair-based structure that is drawn from \texttt{positiveTrain} and \texttt{negativeTrain}. In this setting, the \texttt{<Examples*>} in Listing~\ref{lst:1.1-AmbiguityDetection-FSL-Prompt} is replaced by positive and negative samples depending on the number of shots. A positive sample comprises a feature request and a list of ambiguous text segments extracted from the feature request. These segments are classified as the defined ambiguity sub-class for positive samples, while negative samples indicate ``No Defect Found'' as shown in Listing \ref{lst:1.1-AmbiguityDetection-FSL-Example}. 

We evaluate the extracted ambiguous text segments using six different metrics including: \texttt{Exact Match}, \texttt{Coreff Match}, \texttt{Partial Exact Match}, \texttt{ROUGE-1}, \texttt{ROUGE-2}, \texttt{ROUGE-L}. \texttt{Exact Match} measures whether the predicted ambiguous text segment perfectly matches the Ground Truth (GT) ambiguous text segment. \textit{Coreff Match} accounts for coreferential expressions by considering a predicted argument as a true positive if it matches any of the substrings of the GT. \textit{Partial Exact Match} further relaxes the matching criteria by considering a predicted argument as correct if the GT appears as a substring within it. This ensures that even if the predicted segment is more detailed or slightly re-worded, it is still counted as a valid match. We also incorporate ROUGE-1, which evaluates unigram (individual word) overlap; ROUGE-2, which considers bigram (adjacent word pair) overlap; and ROUGE-L, which assesses the longest common subsequence, allowing for word reordering\cite{akter2022revisiting}, \cite{ESPRE2024}, \cite{morales2024large}. 

\begin{lstlisting}[
  basicstyle=\fontsize{6pt}{8pt}\selectfont,
  lineskip=-0.5pt,
  label = {lst:1.1-AmbiguityDetection-FSL-Prompt},
  caption=Ambiguity Detection using FSL (Without Reasoning)]
{Ambiguity Sub-Class}: Definition  
You are a software analyst specializing in ambiguity detection in GitHub feature requests. 
Carefully read the given statement. Extract and list any text segments containing {Ambiguity Sub-Class} ambiguity from the statement. Multiple segments may contain {Ambiguity Sub-Class} ambiguity; include all of them in a single comma-separated list. Make sure to have all elements of the list in quotation marks. If no segments are found, return No Defect Found. Do not give any explanations, reasoning, or any extra text that is not from the given statement.
Demonstrations are provided for clarity. Each demonstration is separated by the trigger word # END. Inside each demonstration, the statement and extracted segments are separated using the trigger word ####. 
<Examples*>
Statement: <Test Feature Request> 
#### 
Extracted {Ambiguity Sub-Class} segment(s):
\end{lstlisting}

\begin{lstlisting}[
  basicstyle=\fontsize{6pt}{8pt}\selectfont,
  lineskip=-0.5pt,
  label = {lst:1.1-AmbiguityDetection-FSL-Example},
  caption=Example for Ambiguity Detection using FSL (Without Reasoning)]
Statement: <Feature Request> 
#### 
Extracted {Ambiguity Sub-Class} segment(s): [Text segment 1, ... , Text segment n> 
# END 
Statement: <Feature Request> 
#### 
Extracted {Ambiguity Sub-Class} segment(s): No Defect Found
# END 
\end{lstlisting}
\noindent\textbf{Experiment 1.2- Ambiguity Detection using FSL (with Reasoning): } 
To explore the impact of explicit reasoning, we incorporate FSL with reasoning to detect ambiguous text segments for a defined ambiguity sub-class in a feature request. 
This approach first prompts the model to explain why a text segment is ambiguous within its sub-class before presenting the segment. This ordering provides context, improving extraction accuracy and reasoning~\cite{wei2022chain}. We use two settings: Zero-shot learning and FSL with reasoning. The prompt for zero-shot setting includes: definition of the sub-class of ambiguity, persona, instruction, and feature request from the \texttt{testingSet} as shown in Listing \ref{lst:1.2-AmbiguityDetectionWithCoT-FSL-Prompt}, and does not include the demonstration instruction and examples. 

\begin{lstlisting}[
  basicstyle=\fontsize{6pt}{8pt}\selectfont,
  lineskip=-0.5pt,
  label = {lst:1.2-AmbiguityDetectionWithCoT-FSL-Prompt},
  caption=Ambiguity Detection using FSL (With Reasoning)]
{Ambiguity Sub-Class}: {Definition} 
You are a software analyst specializing in ambiguity detection in GitHub feature requests. 
Carefully read the given statement. Extract any text segments containing {Ambiguity Sub-Class} from the statement. Multiple segments may contain {Ambiguity Sub-Class}. If ambiguous segments are found, return a list of tuples where: The first element of each tuple represents the reason why the extracted segment is ambiguous, and the second element is the extracted ambiguous text segment. Strictly make sure that each tuple's elements are enclosed in quotation marks. If no ambiguous segments are found, return "No Defect Found".  
Demonstrations are provided for clarity. Each demonstration is separated by the trigger word # END. Inside each demonstration, the statement and extracted segments are separated using the trigger word ####.  Strictly follow the provided demonstration.  
<Examples*>
Feature Request: <Test Feature Request> 
#### 
Extracted {Ambiguity Sub-Class} segment(s): 
\end{lstlisting}

For FSL with reasoning, our prompt includes: ambiguity sub-class definition, persona, instruction, examples, and a sample for generation from \texttt{testingSet}, as shown in Listing \ref{lst:1.2-AmbiguityDetectionWithCoT-FSL-Prompt}. The \texttt{<Examples*>} are replaced with positive and negative samples from \texttt{positiveTrain} and \texttt{negativeTrain}, depending on the number of shots. A positive sample consists of a feature request paired with a list of tuples.  Each tuple contains two elements: the reasoning behind why a specific text segment is ambiguous and the corresponding extracted text segment. Conversely, 
negative samples indicate ``No Defect Found" as illustrated in Listing \ref{lst:1.2-AmbiguityDetectionWithCoT-FSL-Example}. 
The evaluation follows the approach explained in \textbf{Experiment 1.1}. 

\begin{lstlisting}[
  basicstyle=\fontsize{6pt}{8pt}\selectfont,
  lineskip=-0.5pt,
  label = {lst:1.2-AmbiguityDetectionWithCoT-FSL-Example},
  caption=Example for Ambiguity Detection using FSL (With Reasoning)]
Statement: <Feature Request> 
#### 
Extracted {Ambiguity Sub-Class} segment(s): [('Reason for text segment 1', 'text segment 1'), ... , ('Reason for text segment n', 'text segment n')]
# END 
Statement: <Feature Request> 
#### 
Extracted {Ambiguity Sub-Class} segment(s): No Defect Found
# END 
\end{lstlisting}

\noindent\textbf{Experiment 1.3- Incompleteness Detection: }
For the zero-shot setting, the prompt includes the definition of incompleteness, persona, instruction, and a feature request from the \texttt{testingSet} (Listing~\ref{lst:1.3-IncompletenessDetection-FSL-Prompt}). The model generates a list of missing information or returns ``No Defect Found" if the request is complete. No demonstrations or examples are included in this setting. In the FSL setting, the prompt also includes examples alongside the definition, persona, instruction, and a feature request (Listing~\ref{lst:1.3-IncompletenessDetection-FSL-Prompt}). Each example contains a positive sample (a feature request with missing information) and a negative sample (a complete request) from \textit{positiveTrain} and \textit{negativeTrain}, respectively (Listing~\ref{lst:1.3-IncompletenessDetection-FSL-Example}).

We evaluate incompleteness detection using Precision, Recall, and F1. For the generated missing information, we use Cosine Similarity and manual evaluation. Cosine Similarity includes two scores: \texttt{Complete-List}, which encodes and compares the full predicted and GT lists, and \texttt{Individual-Elements}, which averages the highest similarity scores for each predicted element against the GT. 

\begin{lstlisting}[
  basicstyle=\fontsize{6pt}{8pt}\selectfont,
  lineskip=-0.5pt,
  label = {lst:1.3-IncompletenessDetection-FSL-Prompt},
  caption=Incompleteness Detection]
{Incompleteness}: {Definition} 
You are a software analyst specializing in incompleteness detection in GitHub feature requests. 
Carefully analyze the given feature request statement and determine whether it is incomplete. If the request is incomplete, identify the missing information required for completeness. Include all the missing information in a single comma-separated list. Ensure that every element of the list is enclosed in quotation marks. If the request statement is complete, return Missing Information: No Defect Found. 
Demonstrations are provided for clarity. Each demonstration is separated by the trigger word # END. Inside each demonstration, the statement and missing Information are separated using the trigger word ####. 
<Examples*>
Statement: < Test Feature Request> 
####
Missing Information:  << Generations >> 
\end{lstlisting}

\begin{lstlisting}[
  basicstyle=\fontsize{6pt}{8pt}\selectfont,
  lineskip=-0.5pt,
  label = {lst:1.3-IncompletenessDetection-FSL-Example},
  caption=FSL Example for Incompletensss Detection]
Statement: <Feature Request> 
#### 
Missing Information: ["information1", "information2"] 
# END  
Statement: <Feature Request> 
#### 
Missing Information: No Defect Found
# END 
\end{lstlisting}

\vspace{-1em}
\subsection{Experiment 2: Defect Refinement}\label{sec:Experiment2-DefectRefinement}
To address RQ3 and RQ4, we conduct two ICL experiments using GPT-4o. A positive sample set is created for each ambiguity sub-class and incompleteness. Unlike Experiment 1, each set entails individual defect instances and their corresponding feature requests. 
This allows CQ generation for individual defect instances, focusing on specific ambiguity sub-classes or incompleteness within a feature request. The positive sample set is then split into training and testing sets (i.e., \texttt{positiveTrain} and \texttt{positiveTest}) using a 30/70 split, maximizing the number of instances available while ensuring a sufficiently large test set for evaluation. 

To construct the FSL experiments, we randomly select $n$ samples from \texttt{positiveTrain}. 
Formally, for any \( n \) samples drawn from the training sets: 
$\forall n \in \mathbb{N}, \quad s^+_n \subseteq \texttt{positiveTrain}, \quad |s^+_n| = n$. 
Furthermore, for each experiment with $n$-shot examples, we also consider different orderings, ensuring that different configurations are incorporated in the training process. For any $n$-shot experiment, the total number of valid orderings is given by: $P_n = n!$, where \( P_n \) represents the number of permutations of examples within the experiment.

\noindent\textbf{Experiment 2.1- Ambiguity Refinement: }
We generate CQs for text segments identified as ambiguous for the defined sub-class of ambiguity. Given a feature request, an ambiguous text segment, and the reasoning behind its ambiguity, we generate relevant CQs using two settings: Zero-shot and FSL. 
\noindent\textbf{Experiment 2.2- Incompleteness Refinement: }
To refine an incomplete feature request, we generate CQs based on the reason for incompleteness and missing information from the feature request, using two settings: Zero-shot and FSL settings. 

The prompts for both sub-experiments are published in a readme file on our repository (``Experiment2-DefectRefinement'' under ``ExperimentDesign'' directory)~\cite{AnonymizedRepo2025}. We evaluate generated CQs using (1) cosine similarity\footnote{www.sbert.net/docs/sentence\_transformer/pretrained\_models.html} and (2) manual qualitative evaluation. Cosine similarity measures semantic closeness between generated and GT CQs, reporting two scores: \texttt{Complete List}, which compares the full sets, and \texttt{Individual Elements}, which averages the highest similarity scores for each predicted CQ. 
For the manual evaluation, two authors manually evaluate the correctness of the generated CQs (1) ensuring questions address the identified defects; and (2) assessing differences from GT CQs and reasoning behind them. 

\subsection{Experiment 3: Developer Perspectives}

To explore developers' perspectives on NL defects in feature requests and address RQ5, we conduct interviews with contributors to the Signal Android GitHub repository. We analyze 2,527 closed pull requests, extracting developer profile links and using a web scraper to collect profile details and email addresses. Of 640 unique GitHub usernames, 510 have valid email addresses (excluding those with generic addresses like noreply@github.com). We then send study invitations, with approval from the Institutional Review Board (IRB). Our interview questions follow a structured format, starting with demographic inquiries (see \cite{AnonymizedRepo2025}). The interview is then divided into four main sections, each containing questions designed to address a corresponding overarching interview question (IQ).

\noindent\textbf{IQ1: Developer's Challenges-} IQ1.1. As a developer, what recurring challenges or common ``defects'' do you frequently encounter in feature requests? IQ1.2. How do you perceive ambiguity and incompleteness as challenges in feature requests?

\noindent\textbf{IQ2: Developer's Approach toward Handing Defects-} IQ2.1. What steps do you take when encountering ambiguous or incomplete requests? IQ2.2. What types of questions do you ask to clarify defects? IQ2.3. How has your approach toward addressing defects changed or evolved over time? 

\noindent\textbf{IQ3: Impact of Defects- } IQ3.1. How do defects in requests affect your ability to proceed with your work? IQ3.2. How do you perceive defects as impacting downstream SE tasks (e.g., design, implementation, and testing)? 

\begin{figure*}[h]
\vspace{-1em}
    \centering
    \includegraphics[width=1\textwidth]{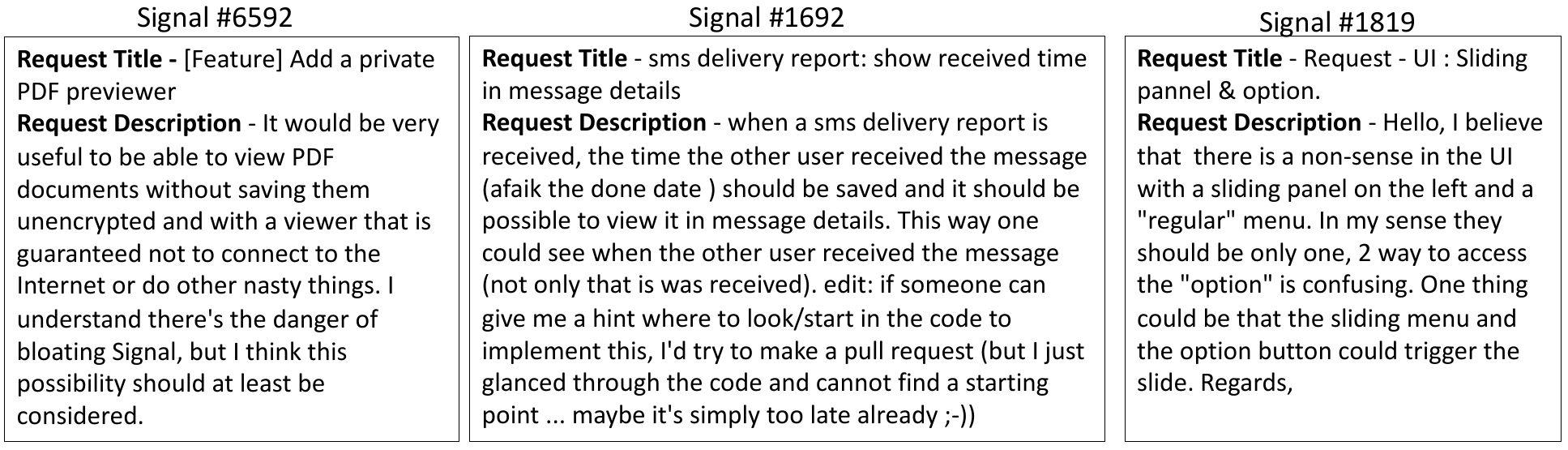}
    \caption{Sampled Feature Requests for the Interviews}
    \label{fig:FeatureRequestForInterview}
\end{figure*}

\noindent\textbf{IQ4: Analysis of Three Feature Requests-} For this part, we randomly select three feature requests (\ref{fig:FeatureRequestForInterview}) from our GT and ask developers to review each feature request. The text of these three feature requests are published online (see~\cite{AnonymizedRepo2025}). Then, we ask three questions that guide the interviewer to identify the defects and CQs. We compare the detected defects and corresponding CQs with the GT. IQ4.1. What are your initial thoughts on this feature request? IQ4.2. If this was an actual feature request assigned to you, what would be your first steps for implementation? IQ4.3. How would you go about clarifying ambiguous or missing parts of the request? 

\section{Evaluation and Results}\label{sec:results}


\subsection{Experiment 1- Results}
\begin{figure}[h]
    \centering
    \includegraphics[width=0.45\textwidth]{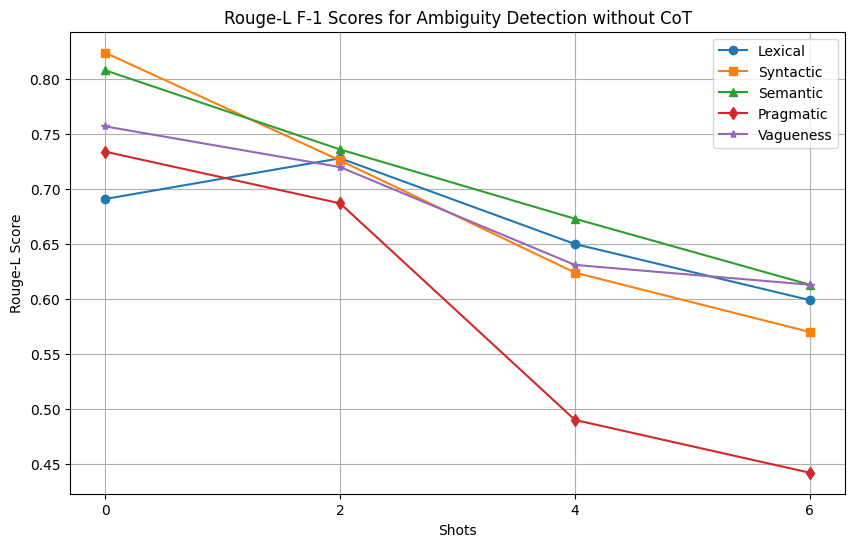}
    \caption{Rouge-L F-1 Scores for Experiment 1.1}
    \label{fig:Experiment1-1-ClassificationWithoutCoT}
\end{figure}

\noindent\textbf{Experiment 1.1- Ambiguity Detection using FSL (Without Reasoning)}: 
Figure~\ref{fig:Experiment1-1-ClassificationWithoutCoT} illustrates the trend in F1 scores for ambiguity detection using FSL (without reasoning) across different shot settings. These F1 scores correspond to the ROUGE-L metric of the best-performing seed in each setting. Notably, the \textbf{zero-shot setting} outperforms all others for four out of five sub-classes of ambiguity, with the exception of the Lexical sub-class, where the highest performance is achieved at the 2-shot setting—followed by a gradual decline in performance as the number of shots increases. We present the detailed results for the zero-shot setting in Table \ref{tab:Experimen1.1-AmbiguityDetectionExerimentWithoutCoT}, which reports F1 scores for five different sub-classes of ambiguities using six evaluation metrics, assessed with three different seeds. Detailed F1 scores for other shot settings are available in our repository\cite{AnonymizedRepo2025}. 

\begin{table*}[]
\vspace{-1em}
\caption{Zero-Shot F1 Scores for Experiment 1.1}
\label{tab:Experimen1.1-AmbiguityDetectionExerimentWithoutCoT}
\resizebox{\textwidth}{!}{%
\begin{tabular}{|c|ccc|ccc|ccc|ccc|ccc|}
\hline
 & \multicolumn{3}{c|}{\textbf{Lexical}} & \multicolumn{3}{c|}{\textbf{Syntactic}} & \multicolumn{3}{c|}{\textbf{Semantic}} & \multicolumn{3}{c|}{\textbf{Pragmatic}} & \multicolumn{3}{c|}{\textbf{Vagueness}} \\ \hline
\textbf{} & \multicolumn{1}{c|}{\textbf{Seed 10}} & \multicolumn{1}{c|}{\textbf{Seed 20}} & \textbf{Seed 45} & \multicolumn{1}{c|}{\textbf{Seed 10}} & \multicolumn{1}{c|}{\textbf{Seed 20}} & \textbf{Seed 45} & \multicolumn{1}{c|}{\textbf{Seed 10}} & \multicolumn{1}{c|}{\textbf{Seed 20}} & \textbf{Seed 45} & \multicolumn{1}{c|}{\textbf{Seed 10}} & \multicolumn{1}{c|}{\textbf{Seed 20}} & \textbf{Seed 45} & \multicolumn{1}{c|}{\textbf{Seed 10}} & \multicolumn{1}{c|}{\textbf{Seed 20}} & \textbf{Seed 45} \\ \hline
\textbf{EM} & \multicolumn{1}{c|}{0.649} & \multicolumn{1}{c|}{0.674} & 0.687 & \multicolumn{1}{c|}{0.824} & \multicolumn{1}{c|}{0.803} & 0.807 & \multicolumn{1}{c|}{0.806} & \multicolumn{1}{c|}{0.804} & 0.806 & \multicolumn{1}{c|}{0.687} & \multicolumn{1}{c|}{0.701} & 0.729 & \multicolumn{1}{c|}{0.744} & \multicolumn{1}{c|}{0.761} & 0.741 \\ \hline
\textbf{Coreff} & \multicolumn{1}{c|}{0.65} & \multicolumn{1}{c|}{0.678} & 0.692 & \multicolumn{1}{c|}{0.824} & \multicolumn{1}{c|}{0.803} & 0.81 & \multicolumn{1}{c|}{0.806} & \multicolumn{1}{c|}{0.804} & 0.806 & \multicolumn{1}{c|}{0.687} & \multicolumn{1}{c|}{0.703} & 0.729 & \multicolumn{1}{c|}{0.744} & \multicolumn{1}{c|}{0.762} & 0.741 \\ \hline
\textbf{Partial Match} & \multicolumn{1}{c|}{0.649} & \multicolumn{1}{c|}{0.674} & 0.688 & \multicolumn{1}{c|}{0.824} & \multicolumn{1}{c|}{0.803} & 0.809 & \multicolumn{1}{c|}{0.807} & \multicolumn{1}{c|}{0.807} & 0.808 & \multicolumn{1}{c|}{0.7} & \multicolumn{1}{c|}{0.711} & 0.74 & \multicolumn{1}{c|}{0.752} & \multicolumn{1}{c|}{0.765} & 0.746 \\ \hline
\textbf{ROUGE-1} & \multicolumn{1}{c|}{0.649} & \multicolumn{1}{c|}{0.679} & 0.692 & \multicolumn{1}{c|}{0.824} & \multicolumn{1}{c|}{0.803} & 0.81 & \multicolumn{1}{c|}{0.808} & \multicolumn{1}{c|}{0.807} & 0.807 & \multicolumn{1}{c|}{0.698} & \multicolumn{1}{c|}{0.714} & 0.735 & \multicolumn{1}{c|}{0.747} & \multicolumn{1}{c|}{0.762} & 0.742 \\ \hline
\textbf{ROUGE-2} & \multicolumn{1}{c|}{0.647} & \multicolumn{1}{c|}{0.667} & 0.68 & \multicolumn{1}{c|}{0.824} & \multicolumn{1}{c|}{0.803} & 0.81 & \multicolumn{1}{c|}{0.806} & \multicolumn{1}{c|}{0.804} & 0.806 & \multicolumn{1}{c|}{0.69} & \multicolumn{1}{c|}{0.704} & 0.73 & \multicolumn{1}{c|}{0.737} & \multicolumn{1}{c|}{0.752} & 0.733 \\ \hline
\textbf{ROUGE-L} & \multicolumn{1}{c|}{0.649} & \multicolumn{1}{c|}{0.677} & 0.691 & \multicolumn{1}{c|}{0.824} & \multicolumn{1}{c|}{0.803} & 0.81 & \multicolumn{1}{c|}{0.808} & \multicolumn{1}{c|}{0.807} & 0.807 & \multicolumn{1}{c|}{0.697} & \multicolumn{1}{c|}{0.713} & 0.734 & \multicolumn{1}{c|}{0.743} & \multicolumn{1}{c|}{0.757} & 0.738 \\ \hline
\end{tabular}%
}
\vspace{-1em}
\end{table*}

\begin{table*}[]
\vspace{-0.25em}
\caption{6-Shot F1 Scores for Experiment 1.2}
\label{tab:Experiment1.2-AmbiguityDetectionCoT}
\resizebox{\textwidth}{!}{%
\begin{tabular}{|c|ccc|ccc|ccc|ccc|ccc|}
\hline
 & \multicolumn{3}{c|}{\textbf{Lexical}} & \multicolumn{3}{c|}{\textbf{Syntactic}} & \multicolumn{3}{c|}{\textbf{Semantic}} & \multicolumn{3}{c|}{\textbf{Pragmatic}} & \multicolumn{3}{c|}{\textbf{Vagueness}} \\ \hline
\textbf{} & \multicolumn{1}{c|}{\textbf{Seed 10}} & \multicolumn{1}{c|}{\textbf{Seed 20}} & \textbf{Seed 45} & \multicolumn{1}{c|}{\textbf{Seed 10}} & \multicolumn{1}{c|}{\textbf{Seed 20}} & \textbf{Seed 45} & \multicolumn{1}{c|}{\textbf{Seed 10}} & \multicolumn{1}{c|}{\textbf{Seed 20}} & \textbf{Seed 45} & \multicolumn{1}{c|}{\textbf{Seed 10}} & \multicolumn{1}{c|}{\textbf{Seed 20}} & \textbf{Seed 45} & \multicolumn{1}{c|}{\textbf{Seed 10}} & \multicolumn{1}{c|}{\textbf{Seed 20}} & \textbf{Seed 45} \\ \hline
\textbf{EM} & \multicolumn{1}{c|}{0.497} & \multicolumn{1}{c|}{0.478} & 0.684 & \multicolumn{1}{c|}{0.585} & \multicolumn{1}{c|}{0.68} & 0.634 & \multicolumn{1}{c|}{0.611} & \multicolumn{1}{c|}{0.47} & 0.622 & \multicolumn{1}{c|}{0.403} & \multicolumn{1}{c|}{0.334} & 0.308 & \multicolumn{1}{c|}{0.402} & \multicolumn{1}{c|}{0.613} & 0.546 \\ \hline
\textbf{Coreff} & \multicolumn{1}{c|}{0.512} & \multicolumn{1}{c|}{0.498} & 0.689 & \multicolumn{1}{c|}{0.586} & \multicolumn{1}{c|}{0.682} & 0.643 & \multicolumn{1}{c|}{0.626} & \multicolumn{1}{c|}{0.479} & 0.624 & \multicolumn{1}{c|}{0.409} & \multicolumn{1}{c|}{0.35} & 0.312 & \multicolumn{1}{c|}{0.428} & \multicolumn{1}{c|}{0.627} & 0.572 \\ \hline
\textbf{Partial Match} & \multicolumn{1}{c|}{0.505} & \multicolumn{1}{c|}{0.48} & 0.685 & \multicolumn{1}{c|}{0.595} & \multicolumn{1}{c|}{0.683} & 0.64 & \multicolumn{1}{c|}{0.611} & \multicolumn{1}{c|}{0.474} & 0.622 & \multicolumn{1}{c|}{0.415} & \multicolumn{1}{c|}{0.336} & 0.332 & \multicolumn{1}{c|}{0.405} & \multicolumn{1}{c|}{0.617} & 0.549 \\ \hline
\textbf{ROUGE-1} & \multicolumn{1}{c|}{0.518} & \multicolumn{1}{c|}{0.496} & 0.688 & \multicolumn{1}{c|}{0.598} & \multicolumn{1}{c|}{0.683} & 0.65 & \multicolumn{1}{c|}{0.625} & \multicolumn{1}{c|}{0.48} & 0.625 & \multicolumn{1}{c|}{0.418} & \multicolumn{1}{c|}{0.342} & 0.323 & \multicolumn{1}{c|}{0.426} & \multicolumn{1}{c|}{0.635} & 0.571 \\ \hline
\textbf{ROUGE-2} & \multicolumn{1}{c|}{0.496} & \multicolumn{1}{c|}{0.476} & 0.682 & \multicolumn{1}{c|}{0.591} & \multicolumn{1}{c|}{0.681} & 0.644 & \multicolumn{1}{c|}{0.611} & \multicolumn{1}{c|}{0.469} & 0.623 & \multicolumn{1}{c|}{0.395} & \multicolumn{1}{c|}{0.305} & 0.308 & \multicolumn{1}{c|}{0.396} & \multicolumn{1}{c|}{0.619} & 0.542 \\ \hline
\textbf{ROUGE-L} & \multicolumn{1}{c|}{0.518} & \multicolumn{1}{c|}{0.496} & 0.688 & \multicolumn{1}{c|}{0.596} & \multicolumn{1}{c|}{0.683} & 0.649 & \multicolumn{1}{c|}{0.625} & \multicolumn{1}{c|}{0.48} & 0.625 & \multicolumn{1}{c|}{0.417} & \multicolumn{1}{c|}{0.341} & 0.323 & \multicolumn{1}{c|}{0.422} & \multicolumn{1}{c|}{0.633} & 0.567 \\ \hline
\end{tabular}%
}
\vspace{-1em}
\end{table*}

\begin{figure}[h]
    \centering
    \vspace{-1em}
    \includegraphics[width=0.45\textwidth]{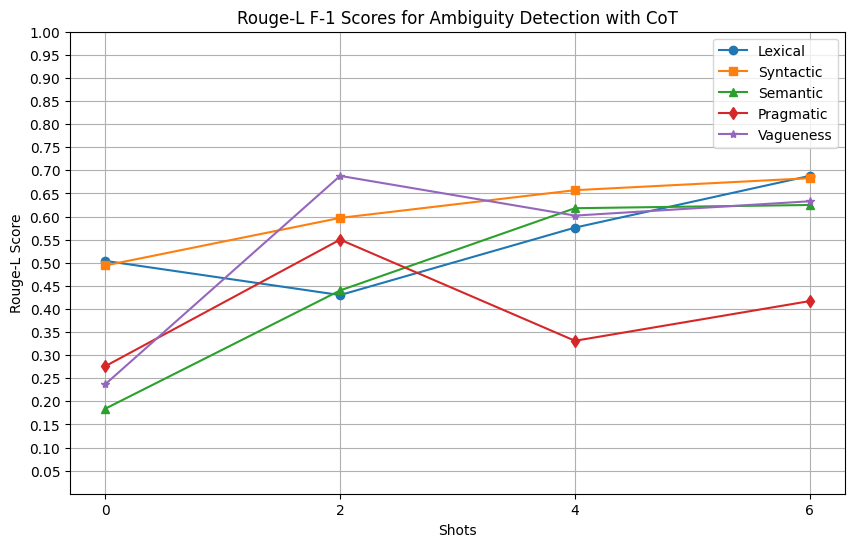}
    \caption{Rouge-L F-1 Scores for Experiment 1.2}
    \label{fig:Experiment1-2-ClassificationWithCoT}
    \vspace{-1em}
\end{figure}

\noindent\textbf{Experiment 1.2- Ambiguity Detection using FSL (With Reasoning)}: 
Figure~\ref{fig:Experiment1-2-ClassificationWithCoT} illustrates the trend in F1 scores for ambiguity detection using FSL (with reasoning) across different shot settings. The scores correspond to the ROUGE-L for the best-performing seed in each shot setting. The \textbf{6-shot setting} yields the highest F1 scores for the Lexical, Syntactic, and Semantic sub-classes. In contrast, the \textbf{2-shot setting} performs best for the Pragmatic and Vagueness sub-classes. We present the detailed results for the 6-shot setting in Table~\ref{tab:Experiment1.2-AmbiguityDetectionCoT}, which reports F1 scores for five different sub-classes of ambiguities using six evaluation metrics, assessed with three different seeds. Detailed F1 scores for zero-, 2-, and 4-shot settings are available in our repository~\cite{AnonymizedRepo2025}. 

\noindent\textbf{Experiment 1.3- Incompleteness Detection}: 
Table~\ref{tab:Experiment1.3-F1ScoreForIncompletenessDetection} presents the precision, recall, and F1 scores for incompleteness detection across different shot settings. The results indicate a slight improvement in precision and F1 score as the number of shots increases, with the 6-shot setting achieving the highest precision and F1 score. However, recall gradually decreases from 1.0 in the zero-shot setting to 0.981 in the 6-shot setting, suggesting that while increasing the number of shots improves precision, it slightly reduces recall. 
Table~\ref{tab:Experimen1.3F1ScoreForIncompletenessDetectionSimilarity} reports Cosine Similarity 
scores across different shot settings, presenting how well the generated missing information aligns with GT. 


\begin{table}[]
\caption{Prec., Rec., and F1 Scores for Experiment 1.3}
\label{tab:Experiment1.3-F1ScoreForIncompletenessDetection}
\begin{tabular}{|p{1.5cm}|p{1.3cm}|p{1.3cm}|p{1.3cm}|p{1.3cm}|}
\hline
\textbf{\# of Shots} & \textbf{0} & \textbf{2} & \textbf{4} & \textbf{6} \\ \hline
\textbf{Precision} & 0.571 & 0.571 & 0.577 & 0.588 \\ \hline
\textbf{Recall} & 1.0 & 1.0 & 0.997 & 0.981 \\ \hline
\textbf{F1} & 0.727 & 0.727 & 0.731 & 0.735 \\ \hline
\end{tabular}
\vspace{-1em}
\end{table}

\begin{table}[]
\caption{Cosine Similarity Score for Experiment 1.3}
\label{tab:Experimen1.3F1ScoreForIncompletenessDetectionSimilarity}
\begin{tabular}{|p{2.7cm}|p{1cm}|p{1cm}|p{1cm}|p{1cm}|}
\hline
\textbf{Similarity Score} & \textbf{0} & \textbf{2} & \textbf{4} & \textbf{6} \\ \hline
\textbf{Complete-List} & 0.142 & 0.223 & 0.192 & 0.210 \\ \hline
\textbf{Individual-Elements} & 0.180 & 0.269 & 0.241 & 0.256 \\ \hline
\end{tabular}
\vspace{-2.5em}
\end{table}


\subsection{Experiment 2- Results} 

\begin{figure}[h]
    \centering
    \includegraphics[width=0.47\textwidth]{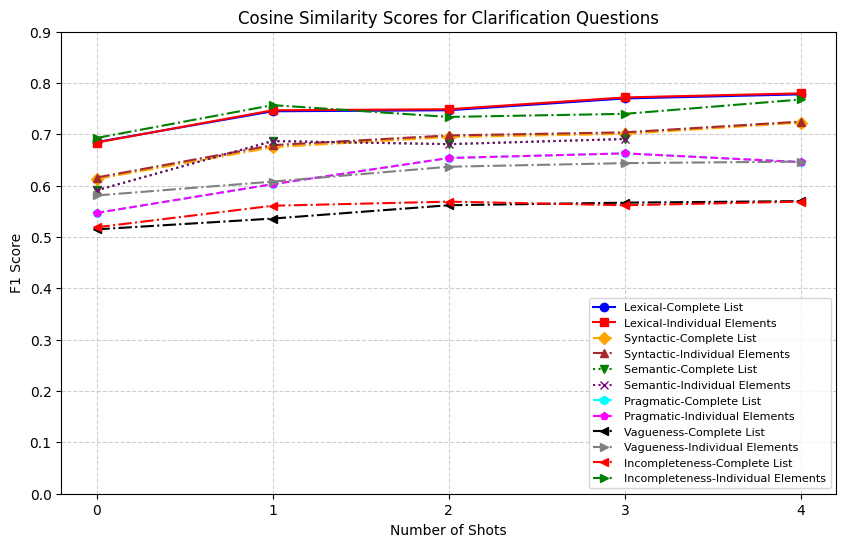}
    \caption{Cosine Similarity Scores for CQs}
    \label{fig:Experiment2-ClarificationQuestions}
    \vspace{-1em}
\end{figure}

Figure~\ref{fig:Experiment2-ClarificationQuestions} presents Cosine Similarity scores for CQ generation across different shot settings for all defect classes. Lexical \& syntactic ambiguity achieve the highest similarity scores at 4-shot. 
Due to limited training data, semantic ambiguity is evaluated only up to 3-shot, where the highest similarity score is observed at 3-shot. 
Pragmatic ambiguity reaches its peak at 3-shot. 
Vagueness and incompleteness attain their highest instance-level F1 score at 4-shot. The GT, generated CQs, and prompts are publicly available online~\cite{AnonymizedRepo2025}. 



To assess the quality of the generated CQs, two authors manually evaluate them based on the criteria mentioned in Section~\ref{sec:Experiment2-DefectRefinement}. Table~\ref{tab:experiment2-results} presents the number of test instances for each defect, along with the count of correctly generated CQs, as determined through this manual evaluation process. 
\begin{table*}[]
    \centering
    \vspace{-2em}
    \caption{Experiment 2- Qualitative Analysis of Generated CQs}
    \label{tab:example}
    \begin{tabular}{|c|c|c|c|c|c|c|}
        \hline
        \multirow{2}{*}{} & \multicolumn{5}{|c|}{\textbf{Ambiguity}} & \multirow{2}{*}{\textbf{Incompleteness}} \\ \cline{2-6} 
        & Lexical & Syntactic & Semantic & Pragmatic & Vagueness &  \\ \hline

        Number of Test Instances & 70 & 17 & 7 & 23 & 10 & 40 \\ \hline
        Number of Correct Generated CQs & 69 & 17 & 6& 22 & 10 & 40 \\ \hline
    \end{tabular}
    \vspace{-2em}
    \label{tab:experiment2-results}
\end{table*}

\subsection{Experiment 3- Results}

We present findings from interviews conducted to address RQ5, which investigates developers' perspectives on NL defects in feature requests. Through our invitations, we receive 15 responses expressing interest, but only seven developers ultimately attend the scheduled interviews.

Each interview session include both the first and last authors as facilitators. The interviewees' ages range from 22 to 41 years old, with geographical distribution as follows: three from the United States, one from Asia, and three from Europe. Their educational backgrounds varies, including high school, undergraduate, and master’s degrees. All participants have been active GitHub contributors/users for over five years. The interviews last an average of 42 minutes and 22 seconds.

\noindent\textbf{IQ1.1. Common Challenges and Defects in Feature Requests:} Below is a summary of the most frequently mentioned challenges, with the number of interviewees citing each issue in parentheses: 
Lack of details (5), Missing information (5), Unclear user intent and goals (2), Lack of prioritization and impact assessment (1), Unknown target audience (1), Misalignment with product mission (requester's familiarity with app objectives) (2), Technical feasibility and implementation complexity (1), Lack of cross-referencing with existing features (1), Scalability, maintainability, and long-term support risks (1), Unclear testing and validation criteria (1), Integration and compatibility issues (1), Lack of continuity and commitment in open-source development (1). 

\noindent\textbf{IQ1.2. Perceptions of Ambiguity and Incompleteness:} 
In general, the interviewees perceive ambiguity and incompleteness as challenges that result in wasted time, inefficiency, multiple ways for implementation, missing implementation details, overlooked edge cases, and unspecified use cases. 
One interviewee notes that these defects often stem from a lack of detail and missing information about the requester’s goal.

\noindent\textbf{IQ2.1. Strategies for Handling Ambiguity and Incompleteness:} 
All interviewees unanimously state that their first step is to ask clarifying questions. Additionally, some participants mention alternative strategies: prototyping, conducting further analysis, and brainstorming solutions. 

\noindent\textbf{IQ2.2. Clarifying Ambiguity and Incompleteness:} 
To gain deeper insights, we ask interviewees to provide examples of the questions they typically ask to resolve ambiguity and incompleteness.
Table~\ref{tab:IQ2.2} summarizes the key questions provided by the interviewees.

\begin{table}[h]
    \centering
    
    \renewcommand{\arraystretch}{1.3}
    \caption{Sample CQs From the Interviews}
    \begin{tabular}{|p{2cm}|p{6.25cm}|} 
        \hline
        \textbf{Topic} & \textbf{Questions} \\
        \hline
        \multirow{5}{=}{\RaggedRight Understanding the Context and Need}  
        & How did this need arise? \\
        & Why do you want this feature? \\
        & What is the problem you are trying to solve? \\
        & What will this feature enable that was not possible before? \\
        & Is there another way to solve the issue? \\
        \hline
        \multirow{4}{=}{\RaggedRight Identifying the Target User and Usage Scenarios}  
        & Who is the user that needs this feature? \\
        & Where do you use this feature? \\
        & What is a specific use case for this feature? \\
        & Can you provide an example on how to use it? \\
        \hline
        \multirow{3}{=}{\RaggedRight Clarifying Expected Behavior and Functional Details}  
        & What should the feature look like, and how should it work? \\
        & What are the expected behaviors in different scenarios (positive and negative cases)? \\
        & What is the typical workflow or process in which this feature is needed? \\
        \hline
        \multirow{3}{=}{\RaggedRight Reproducibility and Technical Validation}  
        & Can you provide steps detailing what you would do? \\
        & Can you show examples, logs, screenshots, or videos to illustrate the issue? \\
        & What platform or environment are you using? \\
        \hline
        \multirow{3}{=}{\RaggedRight Assessing Conflicts and Feasibility}  
        & Does this feature align with the project’s philosophy? \\
        & Are there any potential conflicts or reasons why this approach may not work? \\
        & How does this request fit within ongoing development efforts? \\
        \hline
    \end{tabular}
    
    \label{tab:IQ2.2}
\end{table}

\noindent\textbf{IQ2.3. Evolution of Developers’ Approach Over Time:} 
We ask interviewees whether their approach to handling feature requests had changed over the years. 
Two out of seven interviewees state that they initially began implementation without clarifying the request but have since changed their approach. They now only proceed with implementation once they fully understand the requester’s intent. One interviewee notes that their approach has become more relaxed over time, as they no longer feel pressured to respond or implement feature requests immediately. The remaining four interviewees state that their approach has remained consistent throughout their experience.

\noindent\textbf{IQ3.1. General Impact of Feature Request Defects:} 
To understand the broader implications of ambiguity and incompleteness, we ask interviewees to describe how these defects affect their work. 
The most commonly mentioned impacts include: wasted time and redundant work, decreased motivation and team morale, loss of efficiency and project momentum, misalignment with expectations and overall dissatisfaction, and increased confusion in the development process. 

\noindent\textbf{IQ3.2. Impact on Downstream SE Tasks:} 
Finally, we ask interviewees whether ambiguity and incompleteness affect downstream SE tasks, such as design, implementation, and testing. 
Two interviewees state that the defects have no direct impact on downstream tasks, despite acknowledging general challenges in IQ3.1. Four interviewees highlight testing as the most affected downstream task. Two of those four also mention deployment and maintenance as areas impacted. One interviewee states that these defects influence all downstream SE activities. 


\noindent\textbf{IQ4. Analysis of Three Sample Feature Requests:} We compare the interviewee's analysis of defects in three pre-selected feature requests with our annotations within the GT. The interviewee's answers to questions for the three feature requests are published (see \cite{AnonymizedRepo2025}). Here, we summarize the comparison between the interviewees' analysis and our GT. 




\noindent\textbf{Signal Feature Request \#6592} Seven interviewees analyzed the request for a private PDF viewer in Signal, highlighting ambiguity, incompleteness, feasibility, and security concerns. Comparing their findings with GT annotations reveals key similarities and differences in interpretation. 
Almost all interviewees (i.e., Developers 1, 3, 5, 6, and 7) flagged ``nasty things'' as vague or ambiguous, aligning with our annotation of vagueness. Most interviewees (i.e., Developers 2, 3, 4, 5, and 6) noted that the request lacks critical implementation details, aligning with our classification of incompleteness. Our annotations focused on missing technical implementation details but did not explicitly mention missing UI/UX considerations and user scenarios, which some interviewees highlighted. Developers 1, 2, 3, and 7 raised concerns about whether this feature aligns with Signal’s mission and goals. Our GT annotations did not evaluate feasibility or alignment with Signal’s purpose, focusing only on ambiguity and incompleteness. Further, Developers 3, 5, and 7 emphasized security concerns, with Developer 7 recommending a full security risk analysis. Finally, Developers 3 and 5 suggested researching other messaging apps and GitHub discussions to identify similar feature requests.

\noindent\textbf{Signal Feature Request \#1692} The seven developers analyzed the feature request for adding a received timestamp to SMS delivery reports. Their responses varied, with some developers finding the request clear and within scope, while others highlighted ambiguities, missing UI details, and technical feasibility concerns. Our GT annotations identified two lexical ambiguities (polysemy). Our annotations did not explicitly mention missing UI details, while multiple developers found this to be an important omission.
Our annotations did not assess technical feasibility, while developers questioned whether SMS infrastructure supports this feature. 

\noindent\textbf{Signal Feature Request \#1819} Developers analyzed this request about the ``non-sense in the UI with a sliding panel and a `regular' menu.'' They largely found the feature request confusing, incomplete, and lacking clear feasibility. They focused on missing context, vague terminology, and unclear intent, which aligns with our annotations identifying multiple pragmatic and lexical ambiguities, and incompleteness. 

\section{Discussion}\label{sec:discussion}
To answer RQ1, LLMs can detect and classify defects in feature requests to a considerable extent, though their effectiveness varies across different classes of defects. They perform well in identifying lexical and syntactic ambiguities, showing consistent patterns in detection. However, their ability to handle semantic and pragmatic ambiguities is less reliable, as these often require deeper contextual understanding. While LLMs can recognize vagueness to some extent, their performance may fluctuate depending on the complexity of the request. For incompleteness, LLMs perform well in detecting feature requests as incomplete, consistently achieving high recall. However, their precision is moderate, indicating a tendency to over-predict incompleteness in some cases. To further investigate, we analyze the missing information generated by the model. Among 70 test cases, the model correctly identifies 40 incomplete feature requests and their missing details. Of the remaining 30 requests, 26 are labeled as complete in the GT, but the model classifies them as incomplete, generating missing information. 
Upon review, we find the model’s predictions reasonable, suggesting annotators may have overlooked certain details. The model often highlights privacy concerns and non-functional requirements, which are harder for users to articulate and may require technical expertise. For instance, in Signal Request \#937, the model identifies missing details such as handling sequence numbers, feasibility of vector clocks, timestamp discrepancies, and performance impacts. The remaining four false positives are bug reports misclassified as feature requests, for which annotators did not specify missing information.

To answer RQ2, the prompting method significantly impacts the model’s performance in detecting ambiguity. Initially, FSL without reasoning performs better, particularly in zero-shot setting, but its effectiveness declines as more examples are introduced, leading to a drop in performance with increasing shots. In contrast, FSL when provided reasoning struggles at first, often generating hallucinations due to a lack of structured reasoning, but as more examples are provided, its performance improves significantly, refining its reasoning process and enhancing generalization. Despite these improvements, FSL without reasoning consistently achieves better results, indicating that ambiguity can be effectively detected and classified without explicit reasoning. Given its superior performance, we adopt this setting for detecting incompleteness as well, eliminating the need for adding reasoning in FSL experiments in incompleteness detection.

To answer RQ3, CQs can be generated by leveraging ICL, where the effectiveness of CQ generation varies based on the defect class and the amount of context provided. 
The findings highlight that CQ generation strategies should be tailored to specific ambiguity types, optimizing the number of examples used to maximize effectiveness. 

To answer RQ4, the comparison between LLM-generated and human-generated CQs reveals that LLMs perform exceptionally well in generating accurate CQs across most defect types. 
In many cases, LLM-generated CQs are more detailed and concrete than human-generated ones. 
However, there are cases where the LLM-generated CQ does not match the GT, particularly when a token in the ``Extracted Defective Segment'' appears multiple times in the feature request text. In such instances, the model struggles to determine the correct occurrence of the token, leading to reasoning errors. To address this, additional context could be provided to the model, such as surrounding tokens for the detected segment or indices indicating the exact occurrence of the segment. 

To answer RQ5, developers generally view ambiguity and incompleteness as significant challenges that contribute to wasted time and inefficiencies in the development process. Specifically, developers highlight software testing as the most affected area, emphasizing the difficulty of verifying and validating features when requirements are unclear. Additionally, they extend the impact to deployment and maintenance, suggesting that ambiguous or incomplete requests can lead to uncertainty in implementation and long-term sustainability issues. To address these issues, they consider asking follow-up questions as one of their primary strategies. One of the key differences between the questions posed by developers and those in our GT is their focus and approach. Developers prioritize understanding the requester's goal, often framing their questions as ``Why'' inquiries, which aligns with goal modeling and refinement in RE. In contrast, the questions in our GT primarily aim to disambiguate specific terminology or request additional details to address missing information. While developers also identified these types of questions in their analyses, they consistently began their inquiries by analyzing the request's underlying goal before addressing ambiguity or incompleteness.

Our GT annotations and the developers' analyses align in identifying significant ambiguities within the feature request; however, there are notable differences in how each group approached the request's clarity and actionability. Our annotations focused primarily on lexical and pragmatic ambiguities, highlighting issues such as word choice errors (``nonsense''), polysemy (``slide,'' ``option,'' ``regular menu''), and deictic ambiguity (``on the left''). 
Developers, on the other hand, took a broader, more practical perspective, emphasizing the lack of clarity in the request’s intent and questioning whether there was even a concrete feature to implement. They frequently frame their questions around the ``why'' rather than focusing solely on specific linguistic ambiguities. At the same time, they highlighted challenges such as missing details related to UI aesthetics, interactions, or functionality—concerns that align closely with our observations about incompleteness in feature requests. Several developers noted that the request appeared to be more of an opinion rather than a well-defined feature request. Further, developers consistently requested additional context/information, such as screenshots, videos, or a step-by-step reproduction guide, to better understand the requester’s intent. 
In summary, both groups identified the same problematic terms and phrases, and missing information, reinforcing the validity of our ambiguity and incompleteness analysis. However, our annotations did not assess whether the request was feasible or aligned with the app's goals. In contrast, developers explicitly considered feasibility and the necessity of feature requests in their follow-up clarifications. These findings suggest that future annotation frameworks should incorporate not only ambiguity and incompleteness classification but also the feasibility and goal alignments of feature requests. 

\noindent\textbf{Insights for the RE Community-} 
One key insight is that LLMs can serve as effective assistants during early requirements activities by automatically detecting and classifying common NL defects in feature requests, particularly incompleteness. Their ability to generate clarifying questions tailored to specific ambiguities or missing details offers a valuable starting point for discussion and refinement. This is especially beneficial in large-scale or open-source projects, where manually reviewing every request in detail is often impractical and resource-intensive. Another important observation is the complementary nature of LLMs and human analysts: while developers tend to focus on high-level intent and practical feasibility, LLMs excel at pinpointing localized defects and producing structured follow-up questions. RE tools that integrate these strengths—for example, interfaces that help developers reason about feature goals while receiving LLM-suggested clarifications—can improve the completeness and clarity of requirements. Finally, our results indicate that RE practitioners can adopt LLM-supported workflows not only to streamline defect detection but also to promote more systematic and consistent elicitation practices. For instance, goal-oriented prompting strategies inspired by developer behavior could be embedded into LLMs to produce clarifications that reflect both linguistic cues and stakeholder intent. 
\section{Threats to Validity}\label{sec:threats_validity} 
Internal validity refers to the extent to which our study accurately establishes a causal relationship between feature request defects (i.e., ambiguity and incompleteness) and their impact on detection, CQ generation, and developer perception.


Our dataset includes feature requests from two OSS projects on GitHub, but some bug reports are misclassified as feature requests. While we cannot address this currently, we plan to add bug report classification to our detection model. Additionally, the HTML scraper may introduce structural inconsistencies, which annotators might misinterpret as defects. Since the LLM processes only extracted text, it lacks the original structure. To mitigate this, we aim to incorporate GitHub screenshots and develop models for image-based text analysis to improve accuracy. The accuracy of defect detection and the evaluation of generated CQs depends on human annotations, which may introduce interpretation variability. To mitigate this, we measure inter-annotator reliability, analyzing discrepancies between annotations and ensuring annotators meet regularly to discuss heuristics and refine their approach. 

The performance of LLM-generated CQs may also be influenced by factors unrelated to ambiguity or incompleteness, such as prompt design. To mitigate this, we maintain a consistent prompting approach across experiments, systematically varying only shot settings. To assess consistency, we repeat each prompt 10 times and report the average F1 score in the paper. Additionally, we conduct manual qualitative analysis to ensure the quality of generated results and gain deeper insights into the model’s effectiveness. 

When analyzing developers' perceptions of NL defects, factors such as individual experiences, project background, and familiarity with ambiguity and incompleteness may introduce response bias. While our sample size is limited, it remains diverse in terms of age, geography, education, and experience. We plan to further expand our participant pool by recruiting additional developers and conducting more interviews. 



External validity assesses how well our results generalize beyond our experiments. While tested on two OSS projects, our methods apply to other issue-tracking systems and open-source platforms. The techniques for ambiguity and incompleteness detection and CQ generation can be adapted to various software development environments, extending their relevance beyond the studied repositories.
 
\section{Conclusion and Future work}\label{sec:conclusion}

Overall, we conclude that LLMs effectively detect and classify defects in feature requests, excelling in lexical and syntactic ambiguities while struggling with semantic and pragmatic ambiguities, which require deeper contextual understanding. They also perform well in identifying incompleteness and generating missing information to improve feature requests. Additionally, LLMs prove highly effective in generating clarifying questions, helping refine defective feature requests. Our experiments with GitHub developers validate our approach for detecting and refining NL defects while also revealing new insights into the types of clarifying questions that best aid developers. Furthermore, their feedback highlights new research directions, including feasibility assessment, alignment with project goals and other software requirements, ultimately enhancing the management of software evolution in OSS. 


\section{Acknowledgment}\label{sec:acknowledgment}
The authors thank Dr. Travis Breaux for his suggestions and insights. This work is supported by NSF award \#2318915.

\bibliographystyle{IEEEtran}
\bibliography{reference}

\end{document}